\documentclass{aa}  

\usepackage[pdftex]{color}
\usepackage{txfonts}
\begin{document}

   \title{Impact drag force exerting on a projectile penetrating into a hierarchical granular bed}


   \author{Fumiaki Okubo
          \inst{1}
          \and
          Hiroaki Katsuragi\inst{2}
          }

   \institute{Department of Earth and Environmental Sciences, Nagoya University, Furocho, Chikusa, Nagoya 464-8601, Japan \\
         \and
             Department of Earth and Space Science, Osaka University, 1-1 Machikaneyama-cho, Toyonaka 560-0043, Japan \\
             }

   \date{}

 
  \abstract
   {Impact of a solid object onto a small-body surface can be modeled by the solid impact onto a hierarchically structured granular target.  }
{Impact drag force model for the hierarchically structured granular target is developed based on the experiment. }
   {We perform a set of granular impact experiments in which mechanical strength and porosity of target grains are systematically varied. Tiny glass beads ($5$~$\mu$m in diameter) are agglomerated to form porous grains of $2$--$4$~mm in diameter. Then, the grains are sintered to control their strength. A polyethylene sphere ($12.7$~mm in diameter) is dropped onto a hierarchical granular target consisting of these porous grains. Motion of the penetrating sphere is captured by a high-speed camera and analyzed. }
   {We find that impact drag force produced by the hierarchically structured granular target can be modeled by the sum of inertial drag and depth-proportional drag. The depth-proportional drag in hierarchical granular impact is much greater than that of the usual granular target consisting of rigid grains. The ratio between grain strength and impact dynamic pressure is a key dimensionless parameter to characterize this extraordinary large depth-proportional drag.}
   {Grain fracturing plays an important role in the impact dynamics when the impact dynamic pressure is sufficiently larger than the grain strength. This implies that the effect of grain fracturing should be considered also for the impact on a small body. Perhaps, effective strength of the surface grains can be estimated based on the kinematic observation of the intrusion or touchdown of the planetary explorator.}


   \maketitle
%

\section{Introduction}
\label{sec:introduction}

When a solid projectile impacts on a granular bed, the projectile experiences a drag force which finally halts the projectile motion. Low-speed granular-impact drag force and associated crater formation have been extensively studied in recent decades~\citep{Suarez:2013,Katsuragi:2016,vanderMeer:2017}. First, scaling relations for the penetration depth and crater diameter have been studied~\citep{Walsh:2003,Uehara:2003,Ambroso:2005a}. Then, various related studies relating to cavity formation~\citep{Lohse:2004_2}, numerical modeling to compute stopping time~\citep{Seguin:2009}, splashing dynamics~\citep{Boudet:2006}, and crater formation by soft projectile~\citep{deJong:2017} have been carried out. 

The impact drag force equation has been developed based on the numerical and experimental results~\citep{Tsimring:2005,Ambroso:2005b,Katsuragi:2007,Goldman:2008}. Although the proposed drag force model can reproduce the granular impact dynamics very well, it has some limitations in its applicability. For example, packing fraction of the granular bed significantly affects the impact drag~\citep{Umbanhowar:2010}. Other factors such as container wall~\citep{Seguin:2008,vonKann:2010}, interstitial air~\citep{Royer:2011}, gravity~\citep{Altshuler:2014,AvilaMartinez:2017}, motion history~\citep{Seguin:2019}, and wetness of grains~\citep{Marston:2012}, affect the granular impact dynamics. However, a simple granular impact drag model,
\begin{equation}
  \frac{d^2z}{dt^2} = mg - \frac{mv^2}{d_1} - kz,
  \label{eq:general_drag_eq}  
\end{equation}
has been used as a starting point for the detailed modeling. Here, $m$, $z$, $t$, $g$, $d_1$, and $k$ are mass of projectile, its instantaneous penetration depth, time, gravitational acceleration, and two parameters characterizing the drag force, respectively. In this model, the vertical free-fall impact of a solid projectile onto a granular bed consisting of rigid particles is assumed. The second term in the right-hand side of Eq.~\eqref{eq:general_drag_eq} corresponds to the inertial drag and the third term denotes the depth-proportional drag. The inertial drag usually results from the momentum transfer between projectile and target. This type of inertial drag can be observed even in a usual fluid drag. However, the origin of depth-proportional drag is not very clear. Whereas a simple linear form $kz$ was clearly confirmed in the experiment~\citep{Lohse:2004,Katsuragi:2007}, the scaling of $k$ showed a nontrivial form~\citep{Katsuragi:2013}. Recently, granular Archimedes' law has been considered to explain the depth-proportional drag~\citep{Kang:2018}. In addition, \citet{Roth:2021} revealed that even the well-understood inertial drag cannot be kept constant in the steady deep penetration. However, in the free-fall impact (non-steady) drag, the inertial drag plays a crucial role. In this study, we consider the extension of Eq.~\eqref{eq:general_drag_eq} to the case of hierarchical (porous and fragile) grains to build the firm basis of granular impact dynamics and its planetary application. 

The advantage to use Eq.~\eqref{eq:general_drag_eq} is that it has an analytic solution and scaling laws for material-property dependences of the parameters $d_1$ and $k$. According to \citet{Ambroso:2005b} and \citet{Clark:2013}, this type of equation of motion can be solved in $v-z$ space. Specifically, Eq.~\eqref{eq:general_drag_eq} has been solved as~\citep{Katsuragi:2013},
\begin{equation}
\frac{v^2}{v_0^2} = e^{-\frac{2z}{d_1}}-\frac{k d_1 z}{m v_0^2}
   +\left( \frac{g d_1}{v_0^2}+\frac{k d_1^2}{2 m v_0^2}\right) \left(1- e^{-\frac{2z}{d_1}} \right),
\label{eq:analytic_solution}
\end{equation}
where $v_0$ is the impact velocity defined at the impact moment. In addition, material-property dependences of the parameters $d_1$ and $k$ have also been obtained by the systematic experiments as~\citep{Katsuragi:2013},
\begin{eqnarray}
  \frac{d_1}{D_p} &=& \frac{0.25}{\mu}\frac{\rho_p}{\rho_g} \label{eq:d1_scale}, \\
  \frac{kD_p}{mg} & =& 12 \mu \left( \frac{\rho_g}{\rho_p} \right)^{1/2}, \label{eq:km_scale}
\end{eqnarray}
where $D_p$. $\rho_p$, $\rho_g$, and $\mu$ are the diameter of projectile, density of projectile, bulk density of target granular bed, and its friction coefficient, respectively. Using these relations, we can predict the penetration dynamics of a solid projectile impacting on a granular bed. 

Low-speed granular impact drag is important in planetary science. Because most of the solid bodies in the solar system are covered with granular materials like regolith, granular impact cratering has been studied to understand the cratering mechanics occurring on the planetary surfaces~(see e.g.~\citet{Melosh:1989}). Recently, asteroids have been extensively explored as representative small bodies in the solar system~(e.g.~\citet{Watanabe:2019} and \citet{Lauretta:2019}). To efficiently control the missions of asteroidal surface touchdown and/or sample return, interaction between the probe and granular-regolith surface under the microgravity condition must be properly understood. Because the typical escape velocity is in the order of $10^{-1}$~m~s$^{-1}$ for km-sized asteroids, the impact dynamics in such a low-speed regime should be analyzed. In addition, typical surface gravitational acceleration of such small asteroids is about four orders of magnitude less than that on earth. The effect of gravity might be crucial in the impact cratering dynamics as investigated by \citet{Cintala:1989}. Numerical studies to mimic the explorator situations have been carried out recently~\citep{Ballouz:2021,Sunday:2021,Thuillet:2021}. In these studies, discrete element method (DEM) has been utilized to simulate regolith behaviors. 

Recent observations of asteroid Ryugu suggest that grains covering the astroid have large porosity. This fact was predicted by thermal imaging~\citep{Okada:2020} and confirmed by the returned sample~\citep{Yada:2021}. Such porous grains could be mechanically weak and therefore significantly affect the impact drag force. However, it is difficult to consider grain-level porosity and/or fracturing in DEM simulations. In other words, rigid grains are assumed in usual DEM simulations. Moreover, impact drag measurement using porous grains has not been experimented thus far. 

Impact mechanics among porous dust aggregates have been studied in the context of planetesimal formation~\citep{Blum:2018}. Mechanical characterization and collision outcomes of porous dust aggregates have been experimentally investigated~\citep{Blum:2006,Setoh:2007,Michikami:2007,Guttler:2009,Katsuragi:2017}. Numerical simulations using a porous projectile have also been performed recently~\citep{Planes:2017,Planes:2019,Planes:2020}. However, it is difficult to simulate the behavior of the collection of porous grains due to the computational expense. Only a few number of porous grains (dust aggregates) can be handled in numerical simulations. 

To discuss the impact drag produced by porous grains, their collection has to be investigated. 
The collection of porous grains has hierarchical structure because each porous grain usually consists of numerous tiny monomer particles. Namely, the aggregates consisting of monomer particles hierarchically compose the macroscopic granular matter. Such a hierarchical structure is an interesting research topic both in soft matter physics and planetary science. Hierarchical granular matter is an emergent research field that bridges a gap between soft matter physics and planetary science. Recently, collision of such hierarchical granular clusters have been experimented under the microgravity condition~\citep{Whizin:2017,Katsuragi:2018}. Besides, slow compaction of a hierarchical granular column has also been performed~\citep{Felipe:2021}. However, impact drag force has never been measured in hierarchical granular targets consisting of porous grains while that is quite important to appropriately consider the asteroidal impact phenomena. Therefore, in this study, we perform a simple experiment measuring the impact drag force using porous (fragile) hierarchical granular beds. 
As a result, we find a novel scaling of the drag-force parameter $k$ for the hierarchical granular cases when grain strength is small. In such cases, grain fracturing is the key factor to quantitatively characterize the drag force. In addition, the obtained results are compared with the impact drag by rigid grains or a bulk dust aggregate. 

\section{Experiment}
\label{sec:experiment}
\subsection{Free-fall impact setup}
The experimental system we used in this study is a simple free-fall setup~(Fgi.~\ref{fig:setup}(a)). A polyethylene sphere of diameter $D_p = 12.7$~mm and density $\rho_p = 1,050$~kg~m$^{-3}$ is held by a pull-type electromagnet. This projectile-release unit is mounted on a tall height gauge (Mitsutoyo, HW-100) to control the free-fall height $h$. A granular bed consisting of various kinds of grains (explained in the next subsection) is prepared by simply pouring grains in a cylindrical container with inner radius of $80$~mm and depth of $50$~mm. Any disturbance such as tapping/shaking is not applied during the target preparation. By retracting the movable part of the pull-type electromagnet system, the projectile commences a free fall with zero initial velocity. Then, the dropped projectile impacts on the granular bed. Motion of the projectile is filmed by a high-speed camera (Photron, SA-5). The image-acquiring conditions are as follows: frame rate is $12,000$~fps, size of the image is $896 \times 704$~pixels, and spatial resolution is $30$~$\mu$m~pixel$^{-1}$. The range of free-fall height of the projectile is varied from $h = 10$ to $320$~mm. Thus, the impact velocity ranges $v_0 \simeq 0.44$--$2.5$~m~s$^{-1}$. The identical sphere projectile is used in all experiments. We perform five trials for each experimental condition to confirm the reproducibility. In the following, average data are analyzed unless otherwise noted.

To mimic actual asteroidal situations, microgravity and vacuum environments should be reproduced. However, to reveal the fundamental physical aspect with a simple setup, we performed all the experiments under the atmospheric conditions and 1$g$=9.8~m~s$^{-2}$ gravitational acceleration. In this experiment, only the range of impact speed ($10^{-1}$--$10^0$~m~s$^{-1}$) is close to the asteroidal landing condition.

\subsection{Grains preparation and characterization}
Hierarchical granular beds consisting of porous and fragile grains are prepared by agglomeration and sintering. The first step is to agglomerate monomer grains (tiny glass beads of typical diameter $5~\mu$m ($2$--$10~\mu$m), Potters Ballotini, EMB-10) using a pan-type granulator (AS-ONE, PZ-01R). Monomer grains and small amount of water (2\% of the monomers' mass) are mixed in the rotating pan. Then, the agglomerates are formed due to the effects of capillary bridges and van der Waals force. Subsequently, most of the agglomerates are dried at $105^{\circ}$C for 24 hours by using a drying oven (Yamato Scientific, DVS402) . After that, agglomeraters are sieved to collect the grains of desired size range, $d=2$--$4$~mm. To increase the strength of the agglomerated grains, these grains are sintered. Sintering temperature ($650$, $750$, or $850^{\circ}$C) and duration ($1$, $2$, or $64$~hours) are controlled by using an electric furnace (AS-ONE, SMF-2). To characterize the grains, we measure the friction coefficient $\mu=\tan\theta_\mathrm{r}$ ($\theta_\mathrm{r}$ is angle of repose), bulk density $\rho_g$, bulk (macroscopic) packing fraction $\phi$, and compression strength $Y_\mathrm{g}$. 

\begin{figure}
\begin{center}
    \includegraphics[width=\hsize]{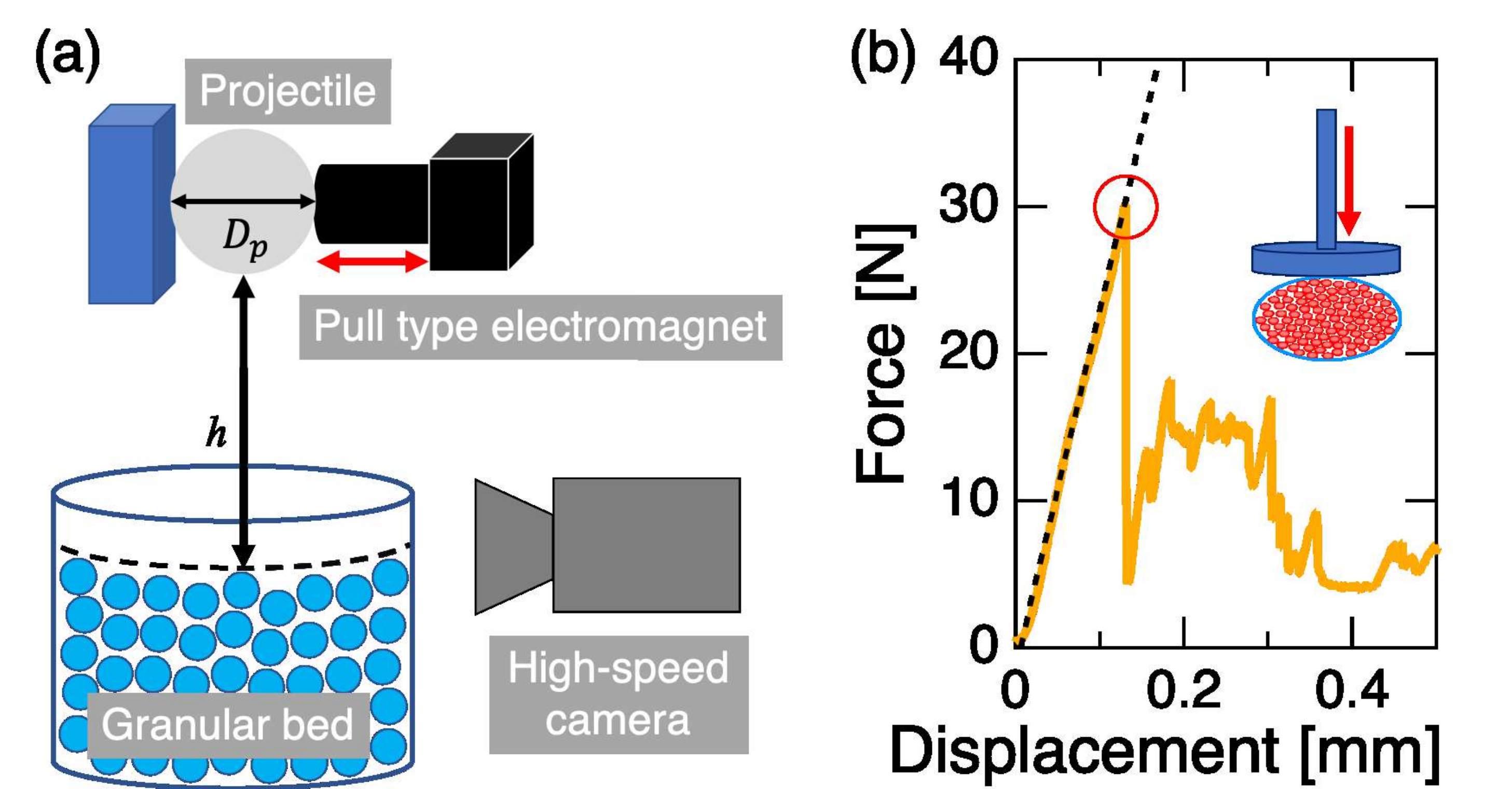}
    \caption{(a)~Experimental setup of the free fall impact of a solid sphere dropped onto a granular target bed. (b)~A typical force curve of the uniaxial compression test of a porous (fragile) grain (S705C01h). In the very early stage of the compression, the compression force linearly increases with displacement (black dashed line). Then, the compression force suddenly decreases at the yielding point (red circle). The peak compression force $F_\mathrm{peak}$ at the yielding point (red circle) divided by the grains crosssectional area is defined as grain strength $Y_\mathrm{g}$. The compression is schematically drawn in the inset.
}
\label{fig:setup}
\end{center}
\end{figure}

To measure $Y_\mathrm{g}$, uniaxial compression tests are performed. A grain is sandwiched by stainless steel plates and vertically compressed by using a universal testing machine (Shimadzu, AG-X). During the compression, applied force and vertical displacement are recorded. 
A typical result of the compression test is shown in Fig.~\ref{fig:setup}(b). 
In the early stage, compression force increases linearly with the displacement. The force curve in this stage is similar to elastic one. When the compression force reaches a certain point, it suddenly drops due to the fracturing. The peak compression force at this point $F_\mathrm{peak}$ is divided by the grains cross-sectional area $A=\pi(d/2)^2=7.1\times 10^{-6}$~m$^2$ ($d=3$~mm) to estimate the strength, $Y_\mathrm{g}=F_\mathrm{peak}/A$. 
In the compression test, compression rate is fixed at 5~mm~min.$^{-1}$. Although this compression rate is much smaller than the free-fall impact speed, we have confirmed that the rate-dependence of the measured strength is limited within several factors over a wide dynamic range of compression rate (although in the slow regime). Due to the technical limitation, we employ $Y_\mathrm{g}$ (measured with slow compression rate) as a typical strengthe value. 

\begin{figure}
\begin{center}
    \includegraphics[width=\hsize]{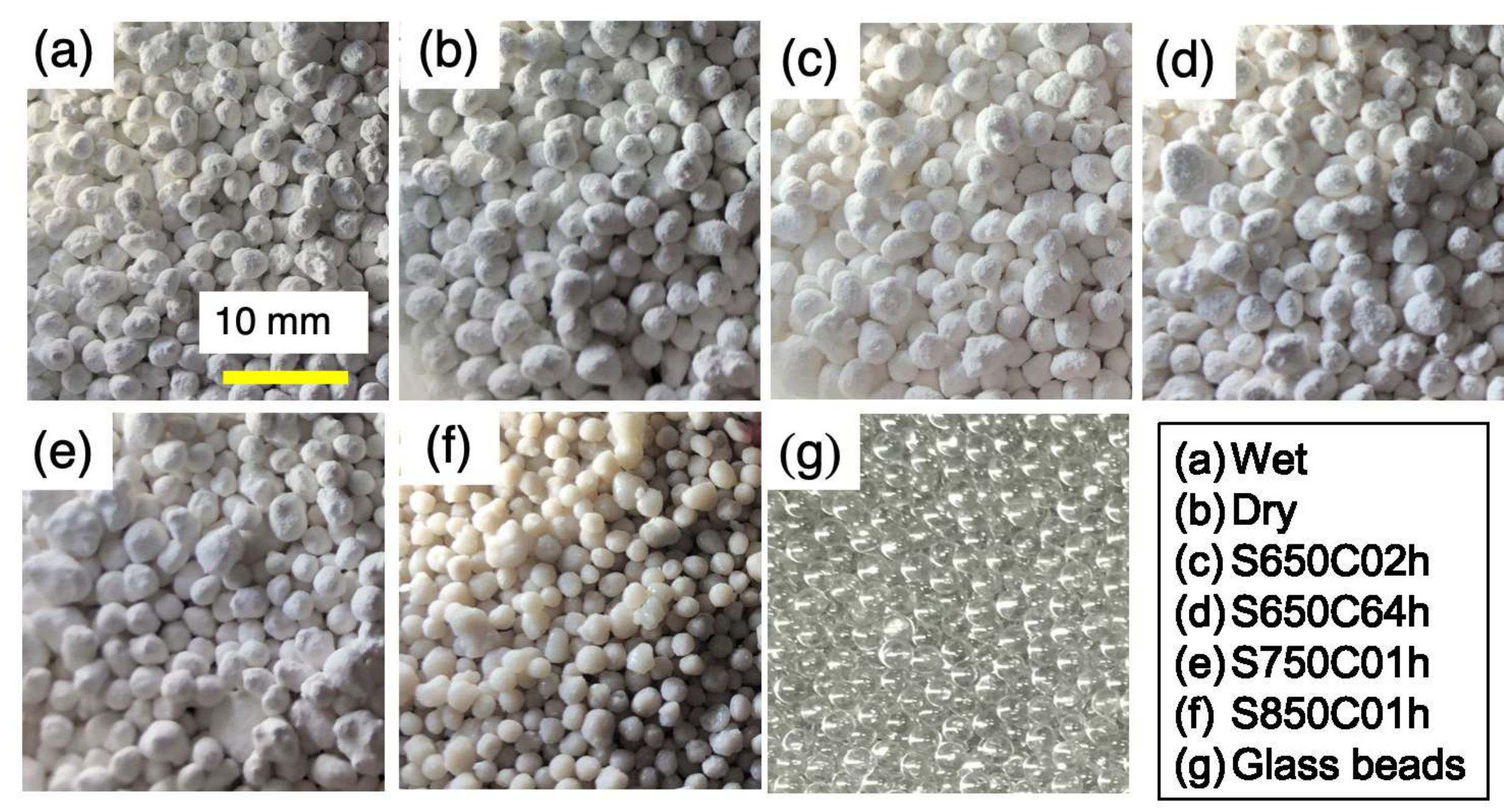}
    \caption{Pictures of (a)-(f) hierarchical grains and (g) rigid grains. The former six hierarchical grains are categorized by its drying or sintering conditions. Each image corresponds to the following granular matter: (a)~wet, (b)~dry, (c)~sintered at $650^{\circ}$C for 2~h (S650C02h), (d) sintered at $650^{\circ}$C for 64~h (S650C64h), (e) sintered at $750^{\circ}$C for 1~h (S750C01h), (f) sintered at $850^{\circ}$C for 1~h (S850C01h), and (g) rigid glass beads. 
}
\label{fig:grains}
\end{center}
\end{figure}

The images and physical properties of the grains used in this experiment are presented in Fig.~\ref{fig:grains} and Table~\ref{tab:grains}, respectively.
The six types of grains (Fig.~\ref{fig:grains}(a-f) and Table~\ref{tab:grains}(a-f)) have hierarchical structure. Among them, four types of grains (Fig.\ref{fig:grains}(c-f) and Table~\ref{tab:grains}(c-f)) are sintered. The grains labeled "S650C02h" are sintered at $650^{\circ}$C for 2 hours. Other labels similarly indicate the sintering temperature and duration. Wet and dry grains are produced only by agglomeration (without sintering). The wet grains are not dried at all while dry grains experience $105^{\circ}$C drying. For comparison, spherical glass beads of 2~mm in diameter are also used as non-deformable (rigid) grains (Fig.~\ref{fig:grains}(g) and Table~\ref{tab:grains}(g)). These granular materials possess the following physical properties: static fraction coefficient $\mu=\tan\theta_r$ ($\theta_r$ is the angle of repose) ranging in $0.45$--$0.88$, bulk density $\rho_g$ ranging in $600$--$1500$~kg~m$^{-3}$, bulk packing fraction $\phi$ ranging in $0.25$--$0.63$, and strength of each grain $Y_\mathrm{g}$ varying in $5.5\times 10^0$--$1.5\times 10^5$~kPa. In general, friction coefficient of hierarchical grains are higher than that of rigid glass beads because of their surface roughness. Moreover, the grains except for [S850C01h] possess quite low bulk density and packing fraction which are consistent with \citet{Katsuragi:2017,Katsuragi:2018}). However, the physical properties of [S850C01h] are almost similar to those of rigid glass beads. This is probably due to the elimination of pores in agglomerated grains, as a consequence of the intense sintering. Anyway, as seen in Table~\ref{tab:grains}, we successfully vary $Y_\mathrm{g}$ over 4 orders of magnitude. 

\begin{table*}
\caption{Physical properties of target grains.}
\centering
\begin{tabular}{lcccc}
\hline\hline
Name & $\mu$ & $\rho_g$ (kg~m$^{-3}$) & $\phi$ & $Y_\mathrm{g}$ (kPa)  \\
\hline
    (a)~Wet & $0.60 \pm 0.02$ & $600 \pm 30$ & $0.25 \pm 0.01$ & $5.5 \pm 1.5$ \\
(b)~Dry & $0.67 \pm 0.03$ & $630 \pm 30$ & $0.26 \pm 0.01$ & $18 \pm 7$ \\
(c)~S650C02h & $0.53 \pm 0.02$ & $660 \pm 30$ & $0.28 \pm 0.01$ & $92 \pm 21$ \\
(d)~S650C64h & $0.73 \pm 0.03$ & $650 \pm 60$ & $0.27 \pm 0.02$ & $(6.0 \pm 3.7)\times 10^2$ \\
(e)~S750C01h & $0.87 \pm 0.03$ & $700 \pm 10$ & $0.29 \pm 0.01$ & $(3.5 \pm 1.5)\times 10^3$ \\
(f)~S850C01h & $0.88 \pm 0.03$ & $1390 \pm 80$ & $0.58 \pm 0.03$ & $(3.9 \pm 1.7)\times 10^4$ \\
(g)~Glass beads & $0.45 \pm 0.02$ & $1500 \pm 60$ & $0.63 \pm 0.03$ & $(1.5 \pm 10.2)\times 10^5$ \\
\hline
\end{tabular}
  \tablefoot{Wet and Dry correspond to grains without sintering. Wet grains are not dried at all. The label S$\mathcal{TTT}$C$\mathcal{XX}$h indicates sintered grains at $\mathcal{TTT}^{\circ}$C for $\mathcal{XX}$~hours. $\mu$, $\rho_g$, and $\phi$ are friction coefficient measured by the angle of repose, bulk density, and packing fraction of the target granular layer, respectively. $Y_\mathrm{g}$ denotes the strength of agglomerated grains. Errors of $\mu$ indicate the measurement uncertainty and other errors indicate the standard deviation of multiple measurements. }
\label{tab:grains}
\end{table*}

\section{Results}
\label{sec:results}
\begin{figure*}
\begin{center}
    \includegraphics[width=\hsize]{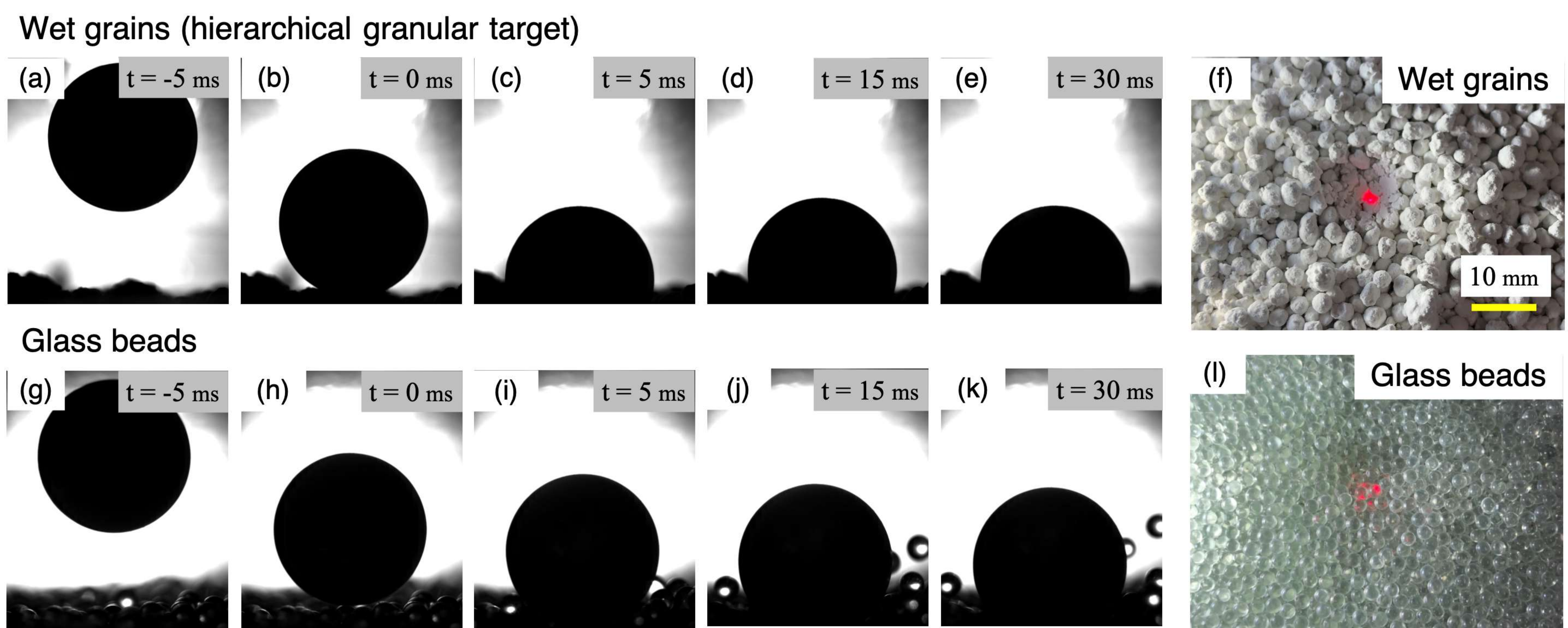}
    \caption{Typical penetration snapshots of the projectile impacting onto granular beds and resultant surface deformation. The panels (a)–(e) show the projectile impacting a hierarchical granular bed (wet grains), and the panels (g)–(k) present the impact onto a rigid glass beads bed. In both situations, the projectile is dropped from $h=80$~mm ($v_0\simeq 1.2$ and $1.4$~m~s$^{-1}$ for wet-grains and glass-beads cases, respectively). In the right panels (f,l), topview images of the impacted surfaces of (f)~hierarchical-granular-bed (wet) case and (l)~rigid-glass-beads case (after removing the projectile) are displayed. The impact points are marked with red laser points.
}
\label{fig:raw_data}
\end{center}
\end{figure*}

Figure~\ref{fig:raw_data} shows typical example images of the projectile penetration and associated surface deformation. The impact moment ($t=0$) at which the projectile bottom reaches the target surface is identified by the video images. Figure~\ref{fig:raw_data}(a-f) shows the results of hierarchical grains (wet type) and Fig.~\ref{fig:raw_data}(g-l) represents the results of rigid glass beads. These examples show the representative behaviors of fragile- and rigid-grains cases, respectively. In both cases, the projectile is dropped from $h = 80$~mm. 
However, penetration behaviors vary depending on the type of target grains. For example, we can observe ejector splashing only in the case of rigid grains (glass beads). In addition, by comparing the surface deformation (Fig.~\ref{fig:raw_data}(f, l)), we find that grains at the impinged zone of the hierarchical granular target are significantly damaged and compressed although the glass beads are not broken at all.

\begin{figure}
\begin{center}
    \includegraphics[width=\hsize]{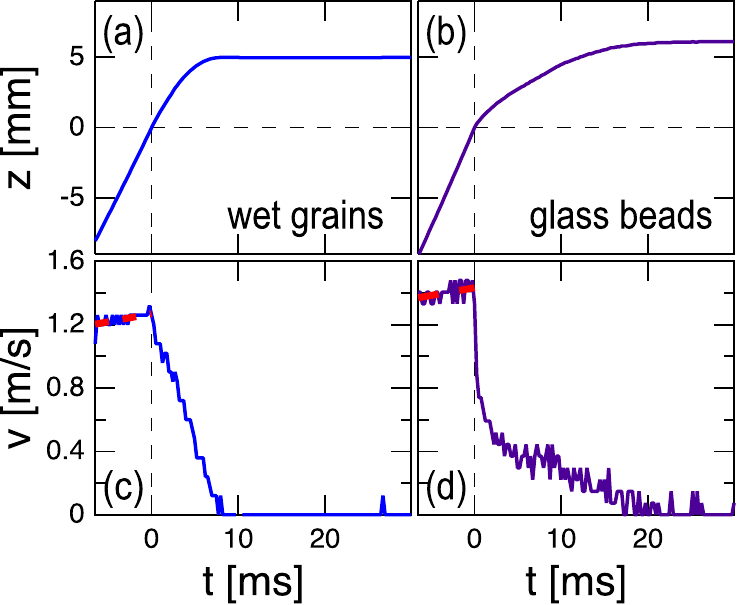}
    \caption{Temporal variations of the penetration depth $z$ and velocity $v$ of the projectile. Panels (a, c) correspond to the case of Fig.~\ref{fig:raw_data}(a-f), and panels (b, d) correspond to the case of Fig.~\ref{fig:raw_data}(g-l). 
  In panels (c, d), gravitational acceleration before the impact can also be confirmed (slope of $g=9.8$~m~s$^{-2}$ is indicated by red dashed lines).
}
\label{fig:zv_t}
\end{center}
\end{figure}

The temporal variation of the projectile position is measured from the raw video data and the example results computed from the data shown in Fig.~\ref{fig:raw_data} are plotted in Fig.~\ref{fig:zv_t}(a, b). 
The level $z=0$ corresponds to the $z$ at the impact moment ($t=0$) and vertically downward direction corresponds to the positive direction of $z$. 
Furthermore, the instantaneous velocity of projectile $v(t)=dz/dt$ can easily be computed as shown in Fig.~\ref{fig:zv_t} (c, d). To reduce the noise level of velocity data, three succesive data points are averaged in the following analysis. As a corollary, projectile velocity at $t<0$ agrees with the constant acceleration with $g=9.8$~m~s$^{-2}$ (red dashed lines in Fig.~\ref{fig:zv_t}(c, d)). The impact velocity is simply defined by $v(t=0)$. After the impact, the projectile impacting hierarchical granular target (wet grains) decelerates relatively in a short time. For the rigid glass beads target, on the other hand, while the early-stage deceleration is significant, the late-stage deceleration is weaker than that in the hierarchical granular target case.  
As a consequence, the maximum penetration depth $z_\mathrm{max}$ becomes larger in the case of glass beads. This is a counterintuitive result because the soft fragile grains causes larger deceleration. These two cases represent the extreme situations of fragile and rigid grains. Thus, we must check the universality of this behavior. 

\begin{figure}
\begin{center}
    \includegraphics[width=\hsize]{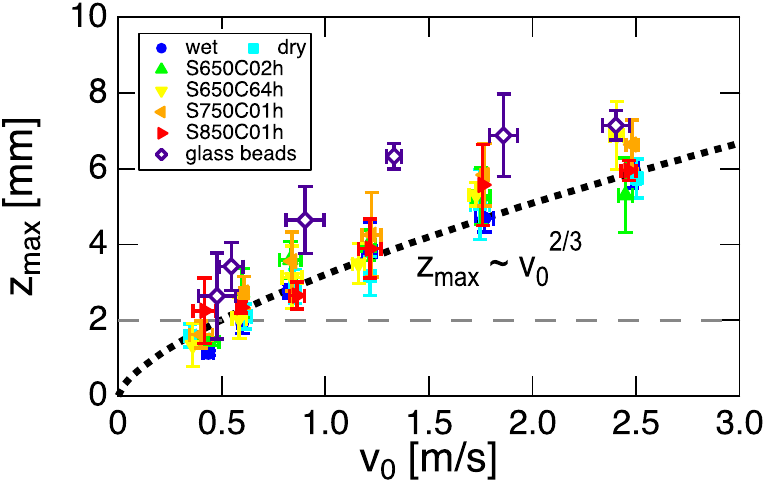}
    \caption{The maximum penetration depth $z_\mathrm{max}$ vs. impact velocity $v_0$ for various granular targets listed in the legend and Table~\ref{tab:grains}. The error bars indicate the standard deviation of five experimental runs. The black dashed line indicates the power-law relation $z_\mathrm{max} \propto v_0^{2/3}$, as a guide to the eye. The data whose $z_\mathrm{max}$ is less than the smallest grain diameter 2~mm (shown by gray dashed line) are not used in the following drag force analysis.
}
\label{fig:z_max}
\end{center}
\end{figure}

To check the generality of this tendency, relations between $z_\mathrm{max}$ and the impact velocity $v_0$ for all types of target grains are plotted in Fig.~\ref{fig:z_max}. The measured data of $v_0$ and $z_\mathrm{max}$ are also listed in Table~\ref{tab:values}. One can confirm that $z_\mathrm{max}$ of hierarchical granular targets is indeed always smaller than that of rigid glass beads. Besides, $z_\mathrm{max}$ is an increasing function of $v_0$ in all cases. Particularly, when $v_0$ is very small ($v_0 \simeq 0.5$~m~s$^{-1}$), $z_\mathrm{max}$ becomes less than $2$~mm. This value ($2$~mm) corresponds to the grain diameter. Therefore, the accurate measurement of $z_\mathrm{max}$ is difficult when $z_\mathrm{max}$ is less than $2$~mm. In addition, the contacting area between projectile and target becomes quite small in this regime. Due to these effects, the very shallow data deviate from the scaling of Eqs.~\eqref{eq:d1_scale} and \eqref{eq:km_scale}, also in the previous studies~\citep{Katsuragi:2013,Katsuragi:2017}. Thus, the data of $z_\mathrm{max}<2$~mm are not used in the following analysis. 
The black dashed curve in Fig.~\ref{fig:z_max} shows the scaling, $z_\mathrm{max} \propto v_0^{2/3}$ which corresponds to $z_\mathrm{max} \propto E_k^{1/3}$ ($E_k$ is impact kinetic energy). This relation has been confirmed in some previous similar experiments~\citep{Uehara:2003,Katsuragi:2017} and qualitatively captures the data trend observed in this study as well. That is, the current experimental result is consistent with the conventional scaling. 

In Fig.~\ref{fig:z_max}, difference in $z_\mathrm{max}$ among fragile grains cannot be clearly observed. The error bars indicating the standard deviation of five experimental runs are too large to clearly distinguish the variation depending on $Y_\mathrm{g}$, particularly in the case of agglomerated fragile grains. This implies that the static quantity such as $z_\mathrm{max}$ is insufficient to characterize the strength-dependent behavior. Therefore, we have to analyze the time-resolved dynamics to quantitatively discuss the penetration dynamics. 

\begin{table*}
  \caption{Measured values of impact velocity $v_0$, penetration depth $z_\mathrm{max}$, and two drag-force parameters $d_1$ and $k$.}
\centering
\begin{tabular}{lcccc}
\hline\hline
Target	&	$v_0$ (m~s$^{-1}$)		&		$z_\mathrm{max}$ (mm)		&		$d_1$ (mm)		&		$k$ (kg~s$^{-2}$)	\\	
\hline
Wet	&	0.44 	$\pm$	0.03 	&	1.14 	$\pm$	0.07 	&	--	&	-- \\
	&	0.60 	$\pm$	0.02 	&	1.93 	$\pm$	0.28 	&	-- 	&	-- \\
	&	0.82 	$\pm$	0.02 	&	2.76 	$\pm$	0.19 	&	4.67 	$\pm$	0.18 	&	50.9 	$\pm$	1.5 \\
	&	1.21 	$\pm$	0.03 	&	3.82 	$\pm$	0.77 	&	6.07 	$\pm$	0.14 	&	25.5 	$\pm$	1.0 \\
	&	1.77 	$\pm$	0.04 	&	4.69 	$\pm$	0.36 	&	8.03 	$\pm$	0.34 	&	76.5 	$\pm$	2.8 \\
	&	2.48 	$\pm$	0.03 	&	6.06 	$\pm$	0.56 	&	9.94 	$\pm$	0.60 	&	87.8 	$\pm$	4.9 \\
\hline
Dry	&	0.35 	$\pm$	0.03 	&	1.59 	$\pm$	0.31 	&	-- 	&	-- \\
	&	0.61 	$\pm$	0.03 	&	2.10 	$\pm$	0.34 	&	4.04 	$\pm$	0.23 	&	51.7 	$\pm$	2.0 \\
	&	0.85 	$\pm$	0.02 	&	2.89 	$\pm$	0.45 	&	4.13 	$\pm$	0.17 	&	47.2 	$\pm$	1.8 \\
	&	1.22 	$\pm$	0.03 	&	3.17 	$\pm$	0.51 	&	4.50 	$\pm$	0.23 	&	69.0 	$\pm$	3.7 \\
	&	1.75 	$\pm$	0.05 	&	5.06 	$\pm$	0.93 	&	6.37 	$\pm$	0.28 	&	51.4 	$\pm$	2.5 \\
	&	2.50 	$\pm$	0.02 	&	5.75 	$\pm$	0.52 	&	8.77 	$\pm$	0.75 	&	89.5 	$\pm$	7.7 \\
\hline
S650C02h	&	0.44 	$\pm$	0.05 	&	1.43 	$\pm$	0.10 	&	-- 	&	-- \\
	&	0.60 	$\pm$	0.01 	&	2.72 	$\pm$	0.64 	&	2.64 	$\pm$	0.10 	&	21.7 	$\pm$	1.0 \\
	&	0.83 	$\pm$	0.05 	&	3.59 	$\pm$	0.48 	&	3.21 	$\pm$	0.08 	&	19.6 	$\pm$	0.8 \\
	&	1.20 	$\pm$	0.03 	&	3.98 	$\pm$	0.40 	&	2.84 	$\pm$	0.09 	&	25.6 	$\pm$	1.7 \\
	&	1.75 	$\pm$	0.04 	&	5.28 	$\pm$	0.74 	&	4.80 	$\pm$	0.23 	&	49.0 	$\pm$	4.4 \\
	&	2.45 	$\pm$	0.03 	&	5.30 	$\pm$	0.99 	&	3.51 	$\pm$	0.14 	&	61.5 	$\pm$	6.7 \\
\hline
S650C64h	&	0.36 	$\pm$	0.01 	&	1.34 	$\pm$	0.57 	&	-- 	&	-- \\
	&	0.58 	$\pm$	0.04 	&	2.05 	$\pm$	0.54 	&	1.29 	$\pm$	0.03 	&	17.1 	$\pm$	0.7 \\
	&	0.84 	$\pm$	0.05 	&	3.13 	$\pm$	0.83 	&	1.74 	$\pm$	0.05 	&	16.1 	$\pm$	1.4 \\
	&	1.16 	$\pm$	0.03 	&	3.49 	$\pm$	0.53 	&	2.13 	$\pm$	0.09 	&	23.7 	$\pm$	2.2 \\
	&	1.73 	$\pm$	0.03 	&	5.32 	$\pm$	0.31 	&	3.30 	$\pm$	0.13 	&	24.3 	$\pm$	1.9 \\
	&	2.41 	$\pm$	0.03 	&	6.88 	$\pm$	0.89 	&	4.01 	$\pm$	0.14 	&	23.4 	$\pm$	2.1 \\
\hline
S750C01h	&	0.40 	$\pm$	0.06 	&	1.61 	$\pm$	0.35 	&	--	&	-- \\
	&	0.61 	$\pm$	0.02 	&	2.73 	$\pm$	0.42 	&	1.33 	$\pm$	0.03 	&	7.66 	$\pm$	0.21 \\
	&	0.84 	$\pm$	0.01 	&	3.55 	$\pm$	0.78 	&	1.62 	$\pm$	0.03 	&	7.04 	$\pm$	0.53 \\
	&	1.22 	$\pm$	0.05 	&	4.25 	$\pm$	1.11 	&	1.78 	$\pm$	0.05 	&	7.33 	$\pm$	0.84 \\
	&	1.77 	$\pm$	0.02 	&	5.84 	$\pm$	0.83 	&	2.54 	$\pm$	0.05 	&	9.80 	$\pm$	1.05 \\
	&	2.48 	$\pm$	0.03 	&	6.64 	$\pm$	0.65 	&	2.65 	$\pm$	0.09 	&	12.7	$\pm$	2.3 \\
\hline
S850C01h	&	0.42 	$\pm$	0.05 	&	2.24 	$\pm$	0.86 	&	1.11 	$\pm$	0.03 	&	7.86 	$\pm$	0.20 \\
	&	0.59 	$\pm$	0.02 	&	2.34 	$\pm$	0.14 	&	1.00 	$\pm$	0.03 	&	7.71 	$\pm$	0.27 \\
	&	0.86 	$\pm$	0.04 	&	2.65 	$\pm$	0.36 	&	0.95 	$\pm$	0.01 	&	6.16 	$\pm$	0.27 \\
	&	1.22 	$\pm$	0.05 	&	3.89 	$\pm$	0.78 	&	1.48 	$\pm$	0.01 	&	5.25 	$\pm$	0.10 \\
	&	1.76 	$\pm$	0.01 	&	5.57 	$\pm$	1.07 	&	1.95 	$\pm$	0.01 	&	3.36 	$\pm$	0.07 \\
	&	2.46 	$\pm$	0.04 	&	5.95 	$\pm$	0.27 	&	1.75 	$\pm$	0.01 	&	2.61 	$\pm$	0.04 \\
\hline
Glass beads	&	0.48 	$\pm$	0.09 	&	2.64 	$\pm$	1.14 	&	1.08 	$\pm$	0.03 	&	4.99 	$\pm$	0.07 \\
	&	0.55 	$\pm$	0.06 	&	3.42 	$\pm$	0.63 	&	1.62 	$\pm$	0.02 	&	4.68 	$\pm$	0.05 \\
	&	0.90 	$\pm$	0.09 	&	4.65 	$\pm$	0.88 	&	2.07 	$\pm$	0.02 	&	4.12 	$\pm$	0.06 \\
	&	1.33 	$\pm$	0.03 	&	6.33 	$\pm$	0.33 	&	2.77 	$\pm$	0.03 	&	3.35 	$\pm$	0.06 \\
	&	1.86 	$\pm$	0.07 	&	6.88 	$\pm$	1.08 	&	2.77 	$\pm$	0.05 	&	3.44 	$\pm$	0.26 \\
	&	2.40 	$\pm$	0.06 	&	7.14 	$\pm$	0.39 	&	3.09 	$\pm$	0.10 	&	2.99 	$\pm$	1.20 \\
\hline
\end{tabular}
  \tablefoot{Errors indicate the standard deviation of five experimental runs. As mentioned in the text, the data with $z_\mathrm{max}<2$~mm are not analyzed.}
\label{tab:values}
\end{table*}

\section{Analysis}
\label{sec:analysis}
\begin{figure}
\begin{center}
    \includegraphics[width=\hsize]{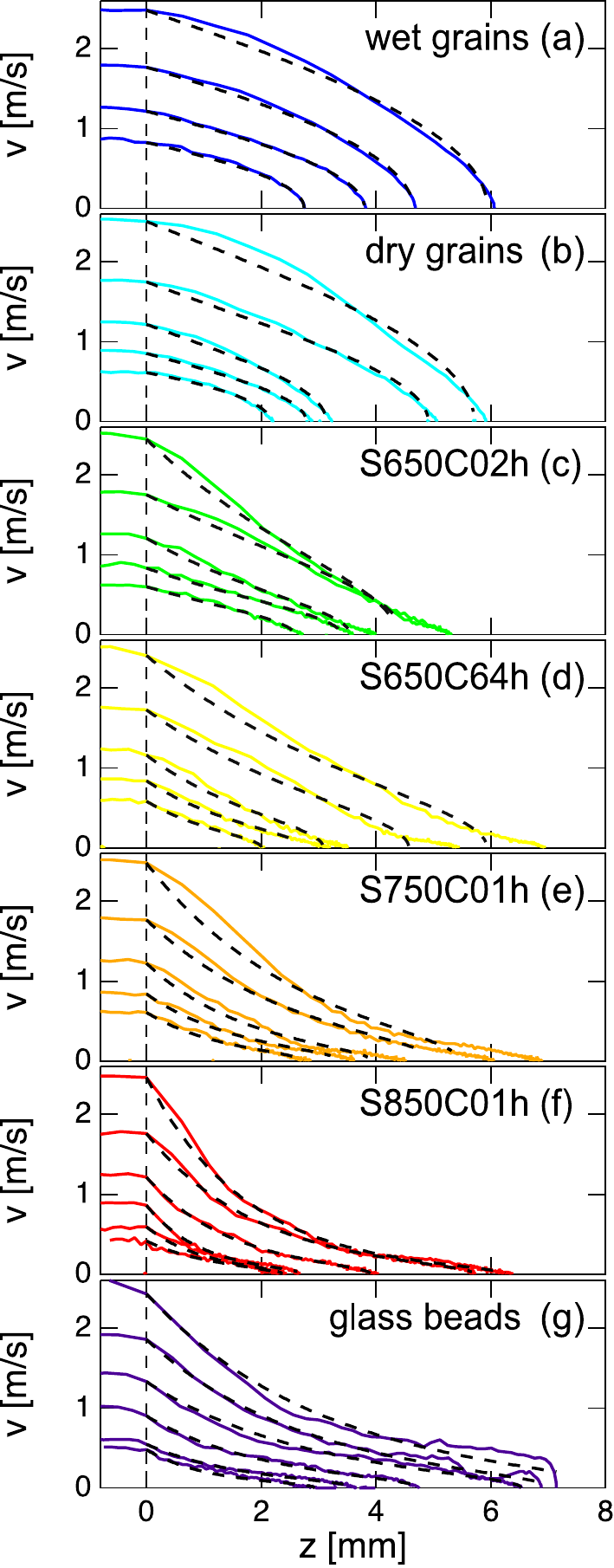}
    \caption{Instantaneous relations between projectile velocity $v$ and penetration depth $z$. The colored solid curves indicate the experimental results. The black dashed curves are the fitting by Eq.~\eqref{eq:analytic_solution}. Although the qualitative behavior of $v(z)$ curves depends on the experimental conditions, all the experimental data can be fitted by the model.  
}
\label{fig:v_z}
\end{center}
\end{figure}

To understand the physical mechanism causing shallow $z_\mathrm{max}$ in hierarchical granular target, quantitative analysis of the impact drag force is necessary. Thus, we consider the applicability of the model of Eqs.~\eqref{eq:general_drag_eq}--\eqref{eq:km_scale} to the hierarchical granular target cases. In Fig.~\ref{fig:v_z}, experimentally obtained $v(z)$ data (colored curves) and the corresponding fittings to Eq.~\eqref{eq:analytic_solution} (black dashed curves) are presented. 
For the hierarchical granular targets consisting of low-strength grains (Fig.~\ref{fig:v_z}(a, b)), gentle deceleration after the impact can be observed as convex shape of the $v(z)$ curve. As the grain strength $Y_\mathrm{g}$ increases, however, the shape of $v(z)$ curves approaches to that of rigid glass beads (Fig.~\ref{fig:v_z}(c-g)). 
The fitting results demonstrate that all the experimental data can be well fitted by the model. This implies that the drag force of hierarchical granular targets can also be expressed by the combination of the inertial drag and the depth-proportional drag, as written in Eq.~\eqref{eq:general_drag_eq}. Meanwhile, variation in the shape of $v(z)$ curves should be explained by the difference in the two fitting parameter values, $d_1$ and $k$.

\begin{figure}
\begin{center}
    \includegraphics[width=\hsize]{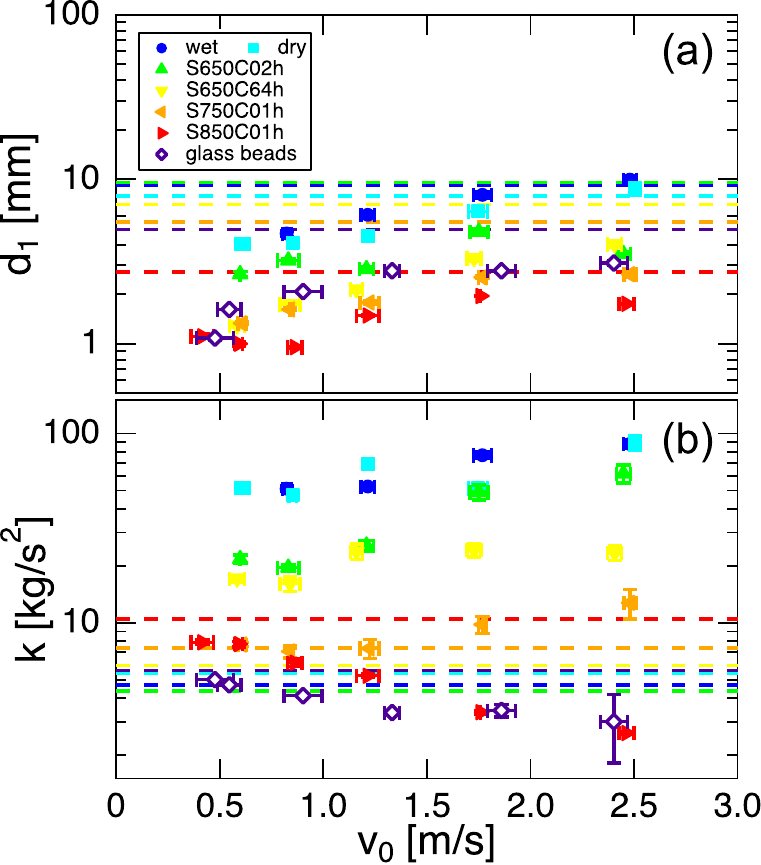}
    \caption{The fitting parameter values: (a)~$d_1$ characterizing the inertial drag force and (b)~$k$ characterizing the depth-proportional drag force. The colored dashed lines in both plots represent the scaling expectations $d_{1s}$ and $k_s$ estimated by Eqs.~\eqref{eq:d1_scale} and \eqref{eq:km_scale}. The deviation of $d_1$ from $d_{1s}$ is less than one order of magnitude and $d_1$ values tend to approach $d_{1s}$ in the large $v_0$ limit. However, $k$ becomes much greater than the scaling expectations particularly in soft (porous and fragile) hierarchical targets. The error bars indicating the standard deviation of five experimental runs.  
}
\label{fig:k_d1}
\end{center}
\end{figure}

\begin{figure}
\begin{center}
    \includegraphics[width=\hsize]{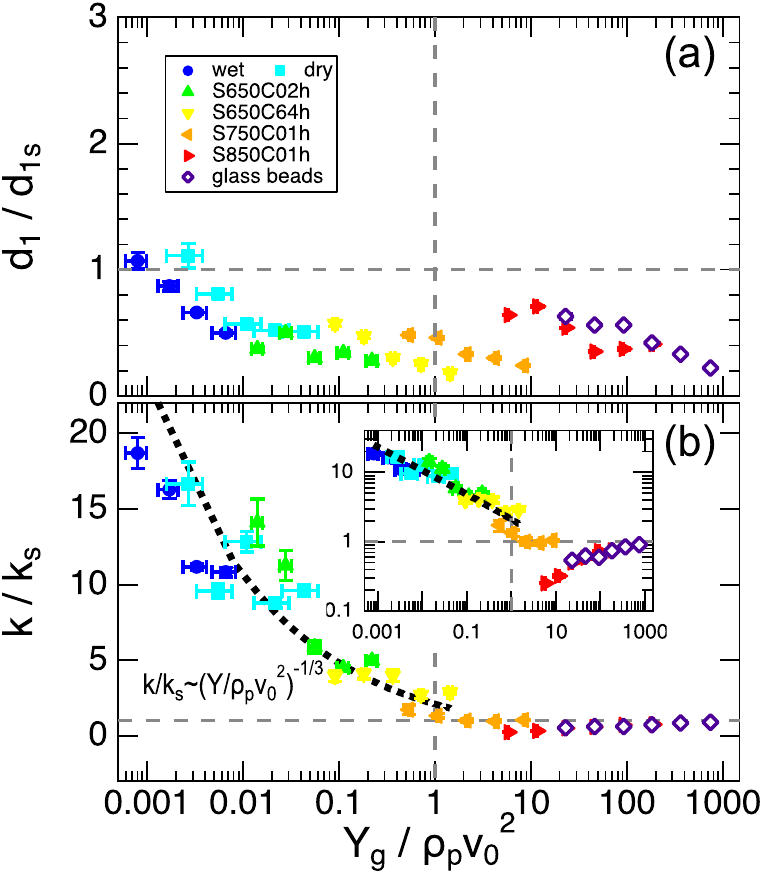}
    \caption{Normalized parameter values vs. the non-dimensionalized grain strength: (a)~$d_1/d_{1s}$ vs. $Y_\mathrm{g}/\rho_p v_0^2$ and (b)~$k/k_s$ vs. $Y_\mathrm{g}/\rho_pv_0^2$. The normalized $d_1$ values distribute around $d_1/d_{1s} \simeq 0.5$. In the regime of $Y_\mathrm{g}/\rho_p v_0^2 \ll 1$, $k/k_s$ can be scaled as $k/k_s=C(Y_\mathrm{g}/\rho_p v_0^2)^{-\alpha}$ with $C=2.4\pm 0.3$ and $\alpha=0.32 \pm 0.03$. In the regime of $Y_\mathrm{g} / \rho_p v_0^2 \gg 1$, $k/k_s$ is close to unity. The inset of (b) shows the log-log plot of the identical data. The error bars indicating the standard deviation of five experimental runs.
}
\label{fig:scaling}
\end{center}
\end{figure}

The estimated values of $d_1$ and $k$ are plotted in Fig.~\ref{fig:k_d1}(a) and (b), respectively. The corresponding data are also listed in Table~\ref{tab:values}. Horizontal dashed lines in Fig.~\ref{fig:k_d1} indicate the expected values computed from the scaling relations for the rigid grains case (Eqs.\eqref{eq:d1_scale} and \eqref{eq:km_scale}). In the following, we refer to the parameter values expected from the scaling of $d_1$ and $k$ as $d_{1s}$ and $k_{s}$, to avoid confusion. Because $\mu$ and $\rho_g$ slightly depend on the target materials (Table~\ref{tab:grains}), the expected values of $d_{1s}$ and $k_s$ distribute around the typical values, $d_{1s} \simeq 5$~mm and $k_s\simeq 5$~kg~s$^{-2}$. Although the range of $d_1$ spans about one order of magnitude (from 1 to 10~mm), the ratio between experimental results and scaling expectations $d_1/d_{1s}$ seem to be almost constant (see also Fig.~\ref{fig:scaling}(a)). This systematic deviation between the scaling expectation $d_{1s}$ and the measured $d_1$ probably comes from the difference in the experimental setup such as system size etc. For instance, the size of container used in this study is not very large compared to the previous studies~\citep{Katsuragi:2007,Katsuragi:2013}. The narrow container might slightly increase the drag force due to the wall effect. Then, the value of $d_1$ becomes small. Besides, velocity dependence of $d_1$ probably results from the very shallow penetration which reduces the actual contacting area between projectile and target. Because these tendencies can be similarly confirmed in all targets, they presumably result from such boundary effects and are not the material-dependent inherent behaviors. Namely, we consider that $d_1$ value is roughly predictable by the scaling of Eq.~\eqref{eq:d1_scale}. Its value is independent of the target grain rigidity. However, the value of $k$ significantly deviates from the expected ones $k_s$ particularly in the small $Y_\mathrm{g}$ regime. The difference exceeds one order of magnitude in the cases of wet and dry grains. 
It can be noticed that hierarchical grains whose $Y_\mathrm{g}$ values are relatively low possess extremely large $k$ value compared with $k_s$. Probably, the hierarchical structure of the grains is responsible for the large value of $k$. Actually, similar large $k$ value has also been confirmed in the impact experiment onto a bulk dust aggregate target~\citep{Katsuragi:2017}. 

Furthermore, $k$ value also depends on $v_0$ (Fig.~\ref{fig:k_d1}(b)). In the hierarchal granular target cases, $k$ shows increasing trend as $v_0$ increases. In rigid grains cases, opposite trend can be confirmed. This means that the effect of $k$ dominates the drag force when $Y_\mathrm{g}$ is small and $v_0$ is large. In other words, the drag force depends not only on the material properties but also on the impact inertia. 

From above observations and analyses, we introduce a dimensionless number $Y_\mathrm{g}/\rho_p v_0^2$ to characterize the degree of grain rigidity. When $Y_\mathrm{g}/\rho_p v_0^2 \gg 1$, grain strength is much greater than the impact inertia (dynamic pressure). Thus, fracturing of grains cannot be induced in this regime. As a result, impact kinetic energy is dissipated by the friction and gravitational potential energy to splash ejector grains. However, when $Y_\mathrm{g}/\rho_p v_0^2 \ll 1$, impact inertia can cause grain fracturing and compression. In this regime, impact kinetic energy is mainly dissipated by grain fracturing. 

Therefore, the value of $k$ could be related to $Y_\mathrm{g}/\rho_p v_0^2$. In Fig.~\ref{fig:scaling}, non-dimensionalized parameters, $d_1/d_{1s}$ and $k/k_s$, are plotted as functions of $Y_\mathrm{g}/\rho_p v_0^2$. As seen in Fig.~\ref{fig:scaling}(a), $d_{1}/d_{1s}$ is roughly independent of $Y_\mathrm{g}/\rho_p v_0^2$. That is why we can consider $d_1$ is almost independent of the grain strength. The weak $v_0$ dependence of $d_1/d_{1s}$ probably originates from the variation in contact area between projectile and target. When $v_0$ is large ($Y_\mathrm{g}/\rho_p v_0^2$ is small), the measured $d_1$ value approaches to $d_{1s}$ because of the sufficient contact area. On the other hand, $k/k_s$ shows non-trivial dependence on $Y_\mathrm{g}/\rho_p v_0^2$. A clear decreasing trend of $k/k_s$ can be confirmed in the regime of $Y_\mathrm{g}/\rho_p v_0^2 \ll 1$. To quantitatively evaluate this trend, data in this regime are fitted to the power-law form,
\begin{equation}
  \frac{k}{k_s} = C \left( \frac{Y_\mathrm{g}}{\rho_p v_0^2} \right)^{-\alpha},
\label{eq:k_ks_scaling}
\end{equation}
where $C=2.4 \pm 0.3$ and $\alpha=0.32 \pm 0.03$ are obtained by the fitting. 
The black dotted curve in Fig.~\ref{fig:scaling}(b) shows the result of this fitting. Ideally, impact-induced grain fracturing should be negligible in the regime of $Y_\mathrm{g}/\rho_p v_0^2 \gg 1$ and the value of $k/k_s$ would converge to $C$ whose value is in the order of unity. Indeed, $k/k_s$ value in $Y_\mathrm{g}/\rho_p v_0^2 \gg 1$ is close to unity. 
Namely, $k$ recovers the conventional scaling value $k \simeq k_s$ when $Y_\mathrm{g}$ is sufficiently greater than $\rho_p v_0^2$. In summary, by introducing a dimensionless number $Y_\mathrm{g}/\rho_p v_0^2$, two limiting behaviors can be seamlessly connected and systematically analyzed by the dimensionless number over six orders of magnitude range.

\section{Discussion}
\label{sec:discussion}
The form of the dimensionless number $Y_\mathrm{g}/\rho_p v_0^2$ is identical to a dimensionless number $\pi_3=Y/\rho_p v_0^2$ defined in the $\Pi$ group scaling for the impact cratering~(e.g.~\citep{Melosh:1989,Katsuragi:2016}). However, they are not the identical quantity. In the conventional $\Pi$ group scaling, $Y$ denotes the bulk strength of the target material. And, $Y_\mathrm{g}$ used in this study is the strength of constituent grains. In general, bulk strength and grains strength are different quantities. In this study, we find that the strength of constituent grains is useful to characterize the impact drag force. In the $\Pi$ group scaling for the impact cratering, $\pi_3$ becomes relevant when the target strength is sufficiently large or the resultant crater size is small enough. This situation corresponds to the typical strength-dominant cratering regime. However, in this study, $Y_\mathrm{g}/\rho_p v_0^2$ becomes a relevant scaling parameter when $Y_\mathrm{g}$ is small or the impact inertia is large. When $Y_\mathrm{g}$ is sufficiently large, we can neglect the effect of grain fracturing and the behavior is dominated by gravity. This tendency is opposite to the conventional crossover between gravity-dominant and strength-dominant regimes in the impact-cratering dynamics. Thus, the scaling found in this study is relevant only when the target material has hierarchical granular structure and the grains internal strength is sufficiently small. 

Substituting the relation of Eq.~\eqref{eq:k_ks_scaling} into Eq.~\eqref{eq:km_scale}, a simple form, 
\begin{equation}
  k = 8C \mu g \sqrt{\rho_p \rho_g} D_p^2 \left( \frac{Y_\mathrm{g}}{\rho_p v_0^2} \right)^{-\alpha}
\label{eq:k_form}
\end{equation}
can be obtained. Note that this relation is applicable only in the regime of $Y_\mathrm{g}/\rho_p v_0^2 \ll 1$. The scaling of original form (Eq.~\eqref{eq:km_scale}) is recovered in $Y_\mathrm{g}/\rho_p v_0^2 \gg 1$. 
Here, we consider the physical meaning of the scaling of Eq.~\eqref{eq:k_form}. This simple scaling can be explained by energy conservation, $E_k = Y_\mathrm{g}V$, where $V$ is the volume of grains fractured by the impact. By carefully observing the impacted hierarchical granular bed, we realize that a certain amount of porous grains are compressed or fractured in the vicinity of impact point. 
By introducing a characteristic length scale of the impact-induced damage zone $D_d$, we simply assume a relation $V \sim D_d^3$. By using projectile density and diameter, the impact kinetic energy is expressed as $E_k \sim \rho_p D_p^3 v_0^2$. Taking them into account, energy conservation can be written as $Y_\mathrm{g} D_d^3 \sim \rho_p D_p^3 v_0^2$. This relation is rewritten as,  
\begin{equation}
  \left( \frac{Y_\mathrm{g}}{\rho_p v_0^2} \right)^{-1/3} \sim  \frac{D_d}{D_p}.
\label{eq:Dd_scaling}
\end{equation}
This power-law form is equivalent to Eq.~\eqref{eq:k_ks_scaling} with $\alpha=1/3$. Substituting Eq.~\eqref{eq:Dd_scaling} into Eq.~\eqref{eq:k_form}, the scaling form of $k$ in the regime of $Y_\mathrm{g}/\rho_p v_0^2 \ll 1$ is expressed by the damaged length scale $D_d$ as follows: 
\begin{equation}
k \sim \mu\sqrt{\rho_p \rho_g} g D_p D_d.
\label{eq:k_Dd}
\end{equation}

This relation is dimensionally sound and includes physically relevant quantities. Based on the actual observation of the impacted targets, we consider $D_d$ is larger than the size of crater produced by the impact. Indeed, wider damage zone has been confirmed in the previous impact experiment using a bulk dust aggregate~\citep{Guttler:2009}. However, Eq.~\eqref{eq:k_Dd} suggests a simple proportionality between $k$ and $D_d$.  
This relation might imply the unreasonably large $D_d$ at the small $Y_\mathrm{g}/\rho_p v_0^2$ limit. In the above discussion, we simply assumed homogeneous energy distribution in the damaged volume $V\sim D_d^3$. However, the degree of damage must be more or less inhomogeneous and localized. To quantitatively discuss the validity of Eq.~\eqref{eq:k_Dd}, we have to measure the compaction and its localization induced by the impact. In other words, Eq.~\eqref{eq:k_Dd} is the first-order approximation and should be improved based on the characterization of the damage zone. The detailed observation of the damage zone by measuring spatial distribution of the fragmented particles and internal porosity is the most important future problem. 

In \citet{Katsuragi:2017}, the unexpectedly large $k$ value was also reported in the impact onto a bulk dust aggregate. In that study, the effective strength of the target material was estimated as $k/\pi D_p = 200$~kPa. Corresponding effective strength for the hierarchical granular target used in this study is $k/\pi D_p \simeq 1$~kPa (a typical value of $k\simeq 40$~kg~s$^{-2}$ is used).   
Namely, the effective strength of a bulk dust aggregate is $10^2$ times greater than that of a hierarchical granular bed. This tendency is qualitatively reasonable because the hierarchical granular matter seems to be weaker than the bulk dust aggregate. According to \citet{Blum:2006} and \citet{Katsuragi:2017}, stress required to open a crack in a bulk dust aggregate (tensile strength) can be estimated as $19$~kPa. This value is close to the strength of the soft hierarchical grains produced in this study ($Y_\mathrm{g}$ value without sintering). The difference in the penetration strength between hierarchical granular bed and bulk dust aggregate could result from the effect of hierarchical grains size. 
Moreover, the actual regolith grains covering planetary bodies have various sizes and strengths. Size and strength distributions might also affect the impact drag force. Such higher-order effects are the future issues to be solved to properly consider the planetary application. The effect of environmental conditions like atmospheric pressure and gravity should also be investigated to discuss the practical planetary application.

In this study, thermal effect is completely neglected. This could be justified only in the low speed regime. When the impact velocity is very large, high temperature produced by the impact could melt the target material. In such situation, the scaling we obtained in this study (Eq.~\eqref{eq:k_ks_scaling}) cannot be applicable. When $Y_\mathrm{g}$ is large, $\rho_p v_0^2$ must also be very large to satisfy $Y_\mathrm{g}/\rho_p v_0^2 \ll 1$. Thus, there must be a certain upper limit of $Y_\mathrm{g}$ above which the completely different physical process dominates the phenomenon. Based on this study, the upper limit of $Y_\mathrm{g}$ is at least greater than $10^3$~kPa because the clear fracturing effect can be confirmed in this ($Y_\mathrm{g}\lesssim 10^3$~kPa) regime. Namely, the impact drag force is dominated by the strength $Y_\mathrm{g}$ only when $Y_\mathrm{g}$ is small. As mentioned above, this is contrastive with the definition of ``strength-dominant impact cratering'' in which the strength $Y_\mathrm{g}$ should be grater than the gravitational pressure $\rho_p g D_p$ (see e.g. \citet{Melosh:1989} or \citet{Katsuragi:2016}). 
Much more systematic studies are necessary for concluding the scaling form including various effects. The parameter variations are still limited in this study. However, we believe that this study opens a new direction of the granular impact study which is a necessary piece also for the planetary science.

Finally, we briefly discuss the possible application potential of this study. For example, if we can record the kinematic information of the penetrator, rober, or any kind of solid object impacting onto a surface of planetary bodies, we can estimate the values of $d_1$ and $k$. Then, if we can independently estimate the density and friction coefficient of the regolith layer, the strength of regolith grains can be computed based on the model developed in this study. In general, strength and density might have a certain relation as seen in Table~\ref{tab:grains}. Once we develop a calibration for that relation, we might be able to estimate various properties of the regolith grains simultaneously. Note that we do not have to perform additional free fall experiment particularly for this analysis. Only we have to do is to record the motion of the instrument which touchdowns the surface of planetary bodies. To properly derive the physical characteristics of the regolith layer, much more systematic experiments measuring damage zone etc. are required as already discussed above. However, this study proposes a new methodology for a simple exploration of fragile (porous) regolith layer by the low-speed impact. 

\section{Conclusions}
\label{sec:conclusions}
In this study, we experimentally investigated the drag force exerting on a projectile impacting onto a bed formed by grains with hierarchical structure. Sintering was used to create hierarchical granular beds with various grain strengths. By the impact, target grains were broken and compressed if target grains are sufficiently weak. From the fittings of $v(z)$ curves of the projectile motion, we found that the impact drag in the hierarchical granular targets can be explained by the drag force model developed for the impact onto a granular bed consisting of rigid grains. However, the contribution of depth-proportional drag force $kz$ increased significantly in the hierarchical granular target case. In this study, we experimentally found a novel scaling relation, $k/k_s \sim (Y_\mathrm{g}/\rho_p v_0^2)^{-1/3}$ at $Y_\mathrm{g}/\rho_p v_0^2 \ll 1$. This implies that the grain fracturing effect dominates the drag force when $Y_\mathrm{g}$ is small compared to $\rho_p v_0^2$. This tendency is in contrast with the conventional $\Pi$ group scaling for impact cratering analysis, in which strength dominates the mechanics when the bulk target strength is relatively large. According to a simple energy conservation assumption, $k$ might relate to the size of damage zone, $D_d$. In addition, we evaluated the penetration strength $k/\pi D_p$ of the target bed directly from the impact drag measurement. As a result, we found that the hierarchical granular bed has an intermediate penetration strength in between a granular bed consisting of rigid grains and a bulk dust aggregate. To obtain the general impact drag-force model which is applicable to various granular targets, we have to consider the effects of other parameters such as thermal effect and grain size in the hierarchical structure.

\begin{acknowledgements}
We thank JSPS KAKENHI for financial support under Grant No.~18H03679.
\end{acknowledgements}

%
   \bibliographystyle{aa} 
   \bibliography{FragileImpact} 
%

\end{document}